\documentclass[manuscript, nonacm]{acmart}

\usepackage{tikz}
\usepackage{tabularx}
\usepackage{array}
\usepackage{hyperref}

\usetikzlibrary{shapes, arrows.meta, positioning, calc, fit, backgrounds}
\AtBeginDocument{%
  }

\setcopyright{acmlicensed}
\copyrightyear{2018}
\acmYear{2018}
\acmDOI{XXXXXXX.XXXXXXX}

\acmConference[Conference acronym 'XX]{Make sure to enter the correct
  conference title from your rights confirmation email}{June 03--05,
  2018}{Woodstock, NY}
\acmISBN{978-1-4503-XXXX-X/2018/06}

\begin{document}


\title{Structure Matters: Evaluating Multi-Agents Orchestration in Generative Therapeutic Chatbots}

\author{Sina Elahimanesh}
\email{siel00002@uni-saarland.de}
\orcid{0000-0001-7251-6661}
\affiliation{%
  \institution{Saarland University}
  \city{Saarbrücken}
  \state{Saarland}
  \country{Germany}
}

\author{Mohammadali Mohammadkhani}
\authornote{These authors contributed equally to this research.}
\email{momo00016@uni-saarland.de}
\affiliation{%
  \institution{Saarland University}
  \city{Saarbrücken}
  \state{Saarland}
  \country{Germany}
}
\affiliation{%
  \institution{Zuse School ELIZA}
  \city{Darmstadt}
  \country{Germany}
}

\author{Sara Zahedi Movahed}
\email{Saza00004@uni-saarland.de}
\authornotemark[1]
\affiliation{%
  \institution{Saarland University}
  \city{Saarbrücken}
  \state{Saarland}
  \country{Germany}
}

\author{Mohammad Mahdi Abootorabi}
\email{mahdi.abootorabi@ece.ubc.ca}
\authornotemark[1]
\affiliation{%
  \institution{The University of British Columbia}
  \city{Vancouver}
  \state{British Columbia}
  \country{Canada}
}

\author{Shayan Salehi}
\email{shayan.salehi81@sharif.edu}
\affiliation{%
  \institution{Sharif University of Technology}
  \city{Tehran}
  \country{Iran}
}

\author{Abbas Edalat}
\email{a.edalat@imperial.ac.uk}
\affiliation{%
  \institution{Imperial College London}
  \city{London}
  \country{United Kingdom}
}

\renewcommand{\shortauthors}{Elahimanesh et al.}

\begin{abstract}

While large language models (LLMs) excel at open-ended dialogue, effective psychotherapy requires structured progression and adherence to clinical protocols, making the design of psychotherapist chatbots challenging. We investigate how different LLM-based designs shape perceived therapeutic dialogue in a chatbot grounded in the Self-Attachment Technique (SAT), a novel self-administered psychotherapy rooted in attachment theory. We compare three architectural variants:
(1) a multi-agent system utilizing finite state machine aligned with therapeutic stages and a shared long-term memory, (2) a single-agent using identical knowledge-base and the same prompts, and (3) an unguided LLM. In an eight-day randomized controlled trial (RCT) with $N=66$ Farsi-speaking participants, balanced across the three chatbots, the multi-agent system is perceived as significantly more natural and human-like than the other variants and achieves higher ratings across most other metrics.
These findings demonstrate that for therapeutic AI, architectural orchestration is as critical as prompt engineering in fostering natural, engaging dialogue.

\end{abstract}


\begin{teaserfigure}
  \centering
  \includegraphics[width=\textwidth]{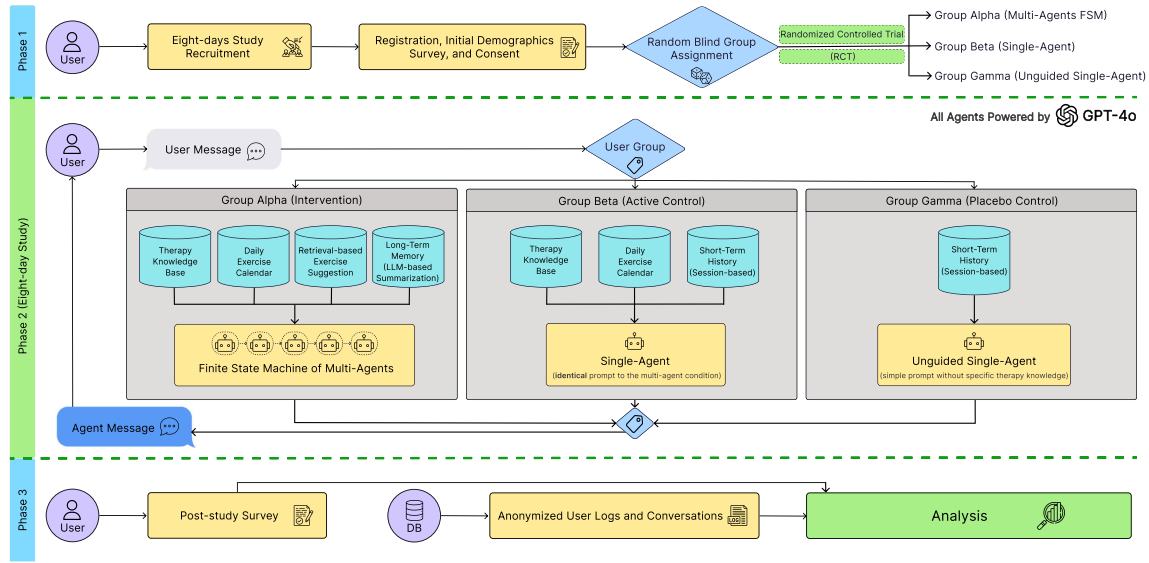}
  \caption{Overview of the user study comprising three phases: (1) recruitment and blinded RCT group assignment; (2) an eight-day study period during which participants interacted with one of three therapeutic chatbot versions, multi-agent FSM-based, single-agent with therapy knowledge, or unguided single-agent; and (3) anonymized post-study survey completion data analysis.}
  \Description{}
  \label{fig:teaser}
\end{teaserfigure}


\maketitle


\section{Introduction}

Conversational agents are increasingly explored as interactive systems for supporting sensitive domains such as mental health, offering on-demand access, anonymity, and reduced cost compared to traditional care \cite{jabbir2023evaluating,shim2025development,lee2025effectiveness,daer2023systematic}. Advances in large language models (LLMs), such as ChatGPT, now enable open-ended, context-sensitive conversations, expanding the expressive capacity of mental health chatbots \cite{10.1145/3392836, arxivGPT4MH2024}. At the same time, structured therapeutic frameworks can be embedded into conversational systems. The Self-Attachment Technique (SAT)~\cite{edalat2015introduction} is a self-administered, attachment-based approach that supports emotion regulation and self-compassion by cultivating an imaginative affectional bond between the adult self and the childhood self of an individual~\cite{alazraki2021empathetic,elahimanesh2023words}. Prior SAT chatbots using rule-based or classification-driven dialogue demonstrated promising empathy and engagement, but were limited in flexibility and scalability \cite{alazraki2021empathetic,elahimanesh2023words}.

Despite these advances, most mental health chatbots emphasize feasibility and acceptability, with comparatively limited attention to the underlying conversational design structure and interaction architecture, leaving open questions about how different architectural paradigms shape user experience ~\cite{shim2025development,lee2025effectiveness,daer2023systematic}. Research highlights the importance of personalization, empathic responses, and sustained engagement \cite{daer2023systematic,jabbir2023evaluating}, yet existing work rarely compares architectural paradigms under controlled conditions \cite{metaagent2025}. Recent proposals such as multi-agent orchestration and finite state machines offer ways to structure complex LLM behavior \cite{metaagent2025}, but their impact on perceived conversational qualities such as naturalness, trust and engagement, particularly for structured interventions remains underexplored.

To address these gaps, we conducted a between-subjects randomized controlled trial (RCT) examining how LLM-based architectural design influences users’ perceived conversational and interaction quality in a SAT-informed chatbot. Using GPT-4o \cite{openai2024gpt4technicalreport} as the base model across all conditions, we compared three architectures. The \textit{Intervention} employed a multi-agent finite state machine inspired by SAT stages, incorporating explicit state transitions, a SAT knowledge base, and shared long-term memory among the agents. The \textit{Active Control} used a single agent with equivalent prompts and SAT knowledge base. The \textit{Placebo Control} used GPT-4o without SAT knowledge or structure, resembling a generic conversational companion. System architecture and prompts were authored in English, while deployment targeted a Farsi-speaking user population.

We deployed the system for eight days with $N=66$ participants (Intervention: 22, Control: 23, Placebo: 21), who completed demographic measures, interacted with their assigned chatbot, and completed post-study surveys. The intervention was perceived as significantly more natural and human-like than both control conditions, with mostly higher but non-significant scores on other self-reported user experience measures, such as trust, memory coherence, user satisfaction, fewer off-topic responses, and ease of use. Chat logs showed the intervention produced more but shorter messages (459 messages, $\sim$230 characters) than the active control (336 messages, $\sim$409 characters) and placebo control (206 messages, $\sim$635 characters), indicating differences in turn-taking patterns and conversational pacing. Overall, this work contributes an empirical comparison of architectural design patterns for LLM-based conversational user interfaces (CUIs), showing how system-level structure shapes conversational interaction and user experience. The evaluation is based on interaction-level metrics from a non-clinical population and does not address clinical outcomes or therapeutic efficacy. These findings suggest implications for how architectural design and memory mechanisms can affect conversational systems for sensitive domains such as mental health.

\section{Related Work}

The Self-Attachment Technique \cite{edalat2015introduction,edalat2017self}, grounded in Bowlby’s attachment theory \cite{bowlby2010separation,bowlby2008loss}, helps users regulate emotional distress by alternating between care-seeking and care-giving roles. Early SAT chatbots operationalize this with rule-based dialogue and emotion classification to provide empathetic responses \cite{alazraki2021empathetic,law2022multilingual,edalat2025self}, enforcing structure but remaining limited in generative flexibility, contextual adaptation, long-term personalization, and multilingual support. Similarly, other therapeutic chatbots, including CBT systems \cite{jabbir2023evaluating,daer2023systematic}, Evebot \cite{yin2019deep}, GPT-2–based therapy bots \cite{wang2021evaluation}, and student distress-detection frameworks \cite{patel2019combating}, combine emotion recognition with neural generation but generally lack explicit therapeutic state modeling, memory, and robustness. The emergence of large language models has enabled richer human–AI collaboration and motivated hybrid neuro-symbolic architectures. Human-in-the-loop systems support cognitive restructuring \cite{sharma2024facilitating}, yet CAPE-II evaluations report failures in crisis handling and boundary maintenance \cite{sobowale2025evaluating}, and excessive trust can amplify harm \cite{song2025typing,choi2025private}. FSM-based systems like MindfulDiary \cite{kim2024mindfuldiary} and ChaCha \cite{seo2024chacha} enforce structured progression while allowing free expression, multi-agent frameworks improve robustness and empathy \cite{metaagent2025,hu2025agentmental}, EmoAgent monitors user state to reduce deterioration \cite{qiu2025emoagent}, and LLMs serve as scalable front-ends for structured intake and assessment \cite{rosteck2025bridging}. Compared to early SAT bots \cite{alazraki2021empathetic,elahimanesh2023words}, these architectures provide stronger guarantees of consistency, safety, and flexibility.

Emotion modeling and personalization are central to therapeutic effectiveness, optimizing empathy, fluency, and novelty \cite{alazraki2021deep}, adapting interventions via real-time mood detection (EMMA; \cite{ghandeharioun2019emma}), and strengthening therapeutic bonds \cite{kocielnik2019will}. RCTs validate their impact: Woebot \cite{fitzpatrick2017delivering} reduced symptoms compared to self-help, while modality and interaction style influence loneliness and emotional dependence \cite{fang2025ai}; other studies highlight emotional resilience \cite{ly2017fully}, humor \cite{sun2024can}, and just-in-time interventions \cite{ghandeharioun2019emma}. Farsi-language systems remain scarce: early Persian SAT chatbots report positive engagement \cite{elahimanesh2023words}, while HamRaz and HamRazEval support culturally grounded therapy \cite{abbasi2025hamraz} and PersianLLaMA provides a foundation for generative modeling \cite{abbasi2023persianllama}, though personalization and fluency remain limited.
Our work synthesizes these strands by embedding a context-aware LLM within a multi-agent, FSM-driven SAT architecture, advancing prior SAT and Persian systems with structured progression, safety-aware generation, long-term engagement, and one of the first RCT-based evaluations of a Farsi SAT chatbot.

\section{Proposed Method}

\textbf{\textit{System Design and Architecture.}}
To investigate how conversational structure and adaptivity influence therapeutic engagement, we developed a three-tier conversational AI system that delivers the Self-Attachment Technique within a \textbf{Randomized Controlled Trial} framework. The study included three experimental conditions: \textbf{(1)} a finite state machine (FSM) driven multi-agent system with long-term memory intervention (Alpha), \textbf{(2)} a single-agent system with a knowledge-base as an active control (Beta), and \textbf{(3)} a minimal single-agent placebo (Gamma). All versions run on a Django REST backend in Python and employ OpenAI's \textit{GPT-4o}~\cite{openai2024gpt4technicalreport} as the conversational engine. User interactions occurred in Farsi, supported by a specialized response retrieval system for intent classification, while prompts and conversational structures were authored in English. To maximize ecological validity, participants accessed the intervention via a React.js web interface, identical across all three conditions. (see Appendix~\ref{appendix:interface} for a screenshot). 


\textbf{\textit{Experimental Conditions and Randomization.}}
Upon registration, participants were randomly assigned to one of three groups and blinded to their condition to minimize expectancy effects. All systems employed the same LLM and shared interface, response style, and interaction modality. 
\textbf{(1) Alpha} implements a 12-state therapeutic protocol based on the SAT stages \cite{elahimanesh2023words}, while supporting open-ended conversation throughout the flow.
(see Appendix~\ref{appendix:fsm} for stage details).
At each stage, dedicated agents guide state transitions using LLM-driven decision logic, while maintaining open-ended dialogue to engage users and smoothly navigate the conversation to the next stage. To ensure consistency among the agents, the system utilizes a shared long-term conversational memory, generated via prompt-guided LLM summarization to preserve emotional states, personal context, and salient experiences. Additionally, it incorporates an agent using an adaptive Retrieval-Augmented Generation (RAG) mechanism to deliver personalized therapeutic exercises;
\textbf{(2) Beta} delivers the identical SAT content and daily exercises but without FSM enforcement, relying solely on a single comprehensive system prompt to implicitly guide the conversation through similar phases within the LLM's context window. To isolate the effect of architecture, this version uses the exact same prompts as Alpha, collapsed into a single sequential prompt;
\textbf{(3) Gamma} provides a minimal conversational single-agent, controlling for the ``digital companion'' effect. It uses a stripped-down prompt with no SAT knowledge, no exercise recommendation engine, and no structured therapeutic goals, representing a near–pure instantiation of the LLM.

\textbf{\textit{Conversational Structure: The Finite State Machine.}}
A key contribution of the Alpha system is its explicit modeling of the therapeutic process. It is operated using a programmatic FSM, based on the SAT flowchart \cite{elahimanesh2023words}.
The FSM encompasses 12 distinct states and divides the interaction into four logical \textbf{Superstates} or phases: \textit{Initiation} (Greeting/Emotion Assessment), \textit{Exploration} (Event Processing), \textit{Intervention} (Exercise Delivery and Feedback Loop), and \textit{Conclusion}.
Unlike rigid rule-based chatbots that rely on keyword matching, the Alpha system employs two advanced transition mechanisms to maintain naturalness: \textbf{(1) LLM-as-Judge Sufficiency Detection} \cite{zheng2023judging, szymanski2025limitations}, where a dedicated prompt evaluates whether accumulated user responses satisfy emotional and informational criteria for the current state, ensuring meaningful expression of the feelings and events before state change; \textbf{(2) Intent-Based Routing}, which classifies over 40 affirmative/negative Farsi linguistic patterns to route the user to the correct next state.

\textbf{\textit{Knowledge Injection and Adaptive RAG.}}
A comprehensive SAT knowledge base, containing theoretical foundations and 27 exercises, was embedded into the system prompts for both Alpha and Beta. Delivery, however, differed: Beta used a static schedule, selecting exercises from a fixed pool based on the agent, while Alpha employed a hierarchical \emph{Adaptive RAG} mechanism. 
This process consists of two phases: \textbf{(1) Constraint-Based Filtering:} The system first retrieves a valid candidate set by filtering the exercise database according to the user’s protocol timeline (Intervention Day 1-8) and therapeutic stage (Beginning, Intermediate, or Advanced). \textbf{(2) Semantic Selection:} A 'Selector' LLM then acts as a re-ranker, analyzing the user’s current emotional state and summarized memory to choose the single most contextually appropriate exercise from the candidate set. It then generates a personalized version of that exercise tailored to the user’s context, minimizing repetition and maximizing personalization.

\textbf{\textit{Longitudinal Memory and Progress Tracking.}}
Alpha maintains structured long-term memory, creating concise summaries every three messages via prompt-guided LLM abstraction to preserve emotional patterns, preferences, and biographical context. A calendar-based model decouples timeline from interaction frequency, ensuring participants resume the correct stage (capped at Day~8) and follow the protocol consistently from beginner to advanced levels.

\section{Experiment and Results}

\textit{\textbf{Participant Recruitment, Demographics, Study Details.}}  
We conducted an eight-day study, following the eight phases of the SAT protocol \cite{edalat2015introduction}, with $N=66$ participants ($43$ male, $23$ female; mean age $27.2$, STD=$10.1$, range $17$-$73$), comparing three system variants: Alpha ($n=22$), Beta ($n=23$) and Gamma ($n=21$). Ethical approval for the user study was obtained from the institution of one of the authors. Participants were recruited by word of mouth and also from Farsi-speaking communities via social media (Twitter, Telegram), participated voluntarily, provided informed consent, and were informed that the system was not a replacement for professional mental health care; logs and survey responses were anonymized. Baseline measures of anhedonia, sadness, anxiety, and uncontrollable worry showed negligible effect sizes ($\eta^2 \leq 0.053$), indicating balanced pre-existing distress. Most participants (47/66, 71.21\%) reported high or very high familiarity with LLMs, while 49 (74.24\%) had no prior familiarity with SAT, suggesting observed effects reflect system design rather than prior knowledge. 
The post-study survey used 5-point Likert items, assessing metrics adapted from prior mental health chatbot evaluations \cite{reguera-gomez-etal-2025-empathy,10919221,info:doi/10.2196/mental.7785,he2023conversational}. Statistical significance was assessed using one-way ANOVA \cite{st1989analysis} with $p$-values from 5{,}000-iteration permutation tests, a nonparametric approach that does not assume normality.

\textbf{\textit{Conversational Naturalness as Primary Effect.}}
The strongest finding was a \textbf{significant architectural effect} on perceived naturalness and human-like tone. Alpha was rated higher ($\mu=3.955$, $SD=0.950$) than Beta ($\mu=3.043$, $SD=0.825$) and Gamma ($\mu=3.211$, $SD=0.787$) ($F=7.017$, $p_{perm}=0.0018$, $\eta^{2}=0.187$),
reflecting a nearly one-point gain on a 5-point scale (0.912 over Beta) and a shift from ``somewhat'' to ``quite'' natural. This corresponds to $\sim$19\% of variance in perceived naturalness, indicating a moderate-to-large effect associated with the Alpha system design. Pairwise tests showed Alpha outperformed both Beta and Gamma, which were similar, indicating that the structured Alpha design was associated with more natural conversational experiences than the comparison conditions. All conditions showed comparable baseline usability (ease of use: $F=0.204$, $p=0.826$, $\eta^{2}=0.007$; feature integration: $F=0.859$, $p=0.446$, $\eta^{2}=0.027$), suggesting that observed differences primarily concern perceived conversational quality rather than functional usability.

\begin{table}[]
\centering
\caption{
Survey results compared three architectures on seven interaction metrics (1–5 Likert). Alpha generally achieved higher mean values than Beta and Gamma across several metrics. For off-topic replies, lower scores are better. Naturalness was statistically significant. F-statistics show variance between versus within groups, and $\eta^2$ indicates the proportion of variance explained by architecture.
}
\label{tab:survey_results}
\begin{tabularx}{\textwidth}{>{\centering\arraybackslash}X
                            >{\centering\arraybackslash}X
                            >{\centering\arraybackslash}X
                            >{\centering\arraybackslash}X
                            >{\centering\arraybackslash}X
                            >{\centering\arraybackslash}X}
\hline
\textbf{Metric} &
\textbf{Alpha} &
\textbf{Beta} &
\textbf{Gamma} &
\textbf{$F$-statistic} &
\textbf{$\eta^2$} \\
\hline
Naturalness*      & \textbf{3.955} & 3.043 & 3.211 & 7.017 & 0.187 \\
Trust             & \textbf{3.136} & 2.739 & 2.737 & 0.737 & 0.024 \\
Empathy           & 3.727 & 3.435 & \textbf{3.895} & 1.357 & 0.043 \\
Memory            & \textbf{3.682} & 3.348 & 3.263 & 1.061 & 0.034 \\
Satisfaction      & \textbf{3.636} & 3.348 & 3.474 & 0.520 & 0.017 \\
Off-topic Replies & \textbf{1.727} & 2.304 & 2.000 & 1.481 & 0.046 \\
Ease of Use       & \textbf{4.000} & 3.826 & 3.789 & 0.204 & 0.007 \\
\hline
\end{tabularx}
\end{table}

\textbf{\textit{Evaluated Interaction Metrics.}}
In this study, we evaluated interaction-level user experience and conversational qualities using self-reported metrics. As shown in Table \ref{tab:survey_results}, Alpha trended higher in the majority of other experiential metrics as well, including memory for past conversations ($\mu=3.682$ vs. 3.348/3.263), perceived empathy ($\mu=3.727$ vs. 3.435/3.895), and trust, while these differences were not statistically significant. Alpha also reported fewer off-topic replies ($\mu=1.727$ vs. 2.304/2.000), indicating better conversational focus, though not significant ($F=1.481$, $p=0.225$, $\eta^{2}=0.046$). Overall satisfaction was moderately positive across groups ($\mu\approx3.4$–3.6, $F=0.520$, $p=0.606$, $\eta^{2}=0.017$), showing general acceptability regardless of architecture. These patterns suggest that architectural design mainly influences perceived naturalness and conversational dynamics, while evidence for effects on relational qualities such as trust and empathy remains inconclusive under the present sample size.


\textbf{\textit{Interaction Dynamics: Why Architecture Matters.}}
To understand \textit{why} the multi-agent system achieved higher perceived naturalness, we analyzed chat logs. One key factor appears to be turn-taking, pacing, and message length. Normalized by group size, Alpha participants generated the most agent messages per person ($\approx 20.9$) versus Beta ($\approx 14.6$) and Gamma ($\approx 9.8$), a 43\% and 113\% increase. Alpha's agent messages were shorter (229.8 characters) than Beta (408.7) and Gamma (635.1), shifting from long, lecture-style responses to rapid, dialogic exchanges that closely emulate human therapeutic conversation. User behavior mirrored this pattern: Alpha participants sent shorter messages on average (29.0 characters) than Beta (38.9) and Gamma (42.8), and the agent-to-user message length ratio was lowest in Alpha (7.9:1) versus Beta (10.5:1) and Gamma (13.4:1), indicating a more balanced, interactive dynamic. 

In addition, importantly the FSM architecture in Alpha, explicitly structured around therapeutic stages, naturally reflects the reasoning of a human therapist, guiding the conversation through problem identification, exploration, and resolution. Multi-agent coordination keeps each agent focused on its stage, while long-term memory preserves context within and across sessions, enhancing continuity and relevance. A structured therapeutic knowledge base ensures contextually appropriate guidance. Beta’s prompting and knowledge contributed to its intermediate performance, and Gamma’s minimal design performed lowest. 
Together, these elements suggest that Alpha’s design, by mimicking human conversational dynamics, multi-agent specialization, memory retention, and grounded expertise, is associated with higher perceived naturalness and improved conversational dynamics. 
\section{Discussion}
Our findings show that architectural choices in LLM-based conversational agents strongly shape perceived interaction quality, beyond prompt design and toward fundamental questions of system structure, due to the staged and process-driven nature of therapy. Multi-agent architectures with explicit state management were associated with more natural interactions than single-agent or unguided LLMs, suggesting that effective CUI design require careful orchestration alongside content delivery.


\textit{\textbf{Conversational Naturalness.}} 
Observed advantages in perceived naturalness reflect more than stylistic preference. Chat logs indicate that the multi-agent system produced shorter, more interactive exchanges, shifting from monologic responses to dialogue-like interactions and challenging the assumption that chatbots should maximize information per turn. Effective CUI design may therefore benefit from mirroring human dialogue rhythms, such as frequent turn-taking and collaborative meaning-making.
By distributing therapeutic stages across specialized agents, the architecture likely enabled this dialogic pattern, with long-term memory further contributing by preserving users’ feelings and experiences across sessions. Notably, gains in naturalness did not extend to empathy, trust, or personalization, suggesting that these are related yet distinct dimensions of the therapeutic alliance. Architectural improvements that enhance naturalness alone may therefore be insufficient to foster emotional attunement, which may depend more strongly on prompt design or response content.


\textit{\textbf{Structural Scaffolding and Design Implications.}}

Across several interaction measures, Alpha generally outperformed one or both comparison conditions, with the clearest and only statistically reliable effect observed for naturalness. This pattern highlights the role of therapeutic structure and architectural scaffolding in shaping perceived interaction quality in CUIs, even without user awareness of system design. Stronger structural guidance supports coherent dialogue, relevant questioning, and alignment with therapeutic goals, without limiting conversational flexibility.
Additionally, these implications can inform the design of conversational agents in other sensitive domains, such medical contexts.

\textit{\textbf{Limitations and Future Directions.}} 
Our study has several limitations and suggests directions for future work:
\textbf{(1)} The eight-day deployment captured only early user impressions, leaving long-term engagement, outcomes, and the development of user-agent relationships unresolved, including whether perceived naturalness translates into sustained use or longer-term benefits.
\textbf{(2)} The use of single-item measures for empathy, trust, and personalization limits construct validity; future studies should employ validated multi-item instruments (e.g., the Working Alliance Inventory \cite{hatcher2006development}) and examine how architectural effects vary across user characteristics.
\textbf{(3)} The focus on Persian-speaking, educated participants (93.9\% holding at least a Bachelor’s degree) constrains cross-linguistic and broader population generalizability. Future work could explore more diverse linguistic, cultural, and educational populations.
\textbf{(4)} No clinical or therapeutic outcomes were assessed, and findings reflect perceived interaction quality rather than efficacy. Future work can investigate the effect of interaction metrics on therapy efficiency in a clinical population.
\textbf{(5)} The bundled design (multi-agent orchestration and memory) prevents attributing effects to specific components. Because in the multi-agent variation there must be mechanisms for information sharing among agents, it was not possible to implement it without a shared memory. Thus, in the intervention version, multi-agent architecture and memory coexist, which represents a limitation of our work. While we preserved exactly equal prompts and knowledge bases for the alpha and beta versions, the findings indicate that multi-agent architecture combined with memory and knowledge base performed better compared to a single-agent system with the same knowledge base, and also outperformed a general unguided LLM agent. Future work can investigate each of these components separately through an ablation study.
\textbf{(6)} The collected chatbot interaction data represent a valuable therapeutic dataset that, when combined with other sources, could support LLM alignment using methods such as PPO or DPO \cite{zheng2023secretsrlhflargelanguage, rafailov2024directpreferenceoptimizationlanguage}.

\section{Conclusion}

Architectural design strongly shaped user experience in our therapeutic LLM chatbot. Multi-agent, memory-augmented systems with explicit state management produced more natural, human-like interactions than single-agent or unguided alternatives, with higher (though non-significant) relational ratings. By holding content constant, we isolated structural effects often conflated with content. These results suggest that advancing therapeutic AI requires not only capable LLMs and refined prompts, but also careful architectural orchestration to support structured therapeutic progression.



\bibliographystyle{ACM-Reference-Format}
\bibliography{references}

\newpage 
\appendix

\section*{Appendix}

\section{Finite State Machine Specification}
\label{appendix:fsm}

The Alpha (Intervention) system is operated using a programmatic Finite State Machine, comprising 12 distinct states, based on the flowchart of the SAT \cite{elahimanesh2023words} algorithm. This architecture enforces the therapeutic structure of the Self-Attachment Technique, while maintaining conversational flexibility through LLM decision nodes. The FSM is divided into four logical phases: Initiation, Exploration, Intervention, and Conclusion.





\subsubsection{Phase I: Initiation and Assessment}
\begin{itemize}
    \item \textbf{State 1: GREETING\_FORMALITY\_NAME}
    \textit{Objective:} Establish rapport and configure user preferences.
    \textit{Logic:} The agent greets the user and optionally requests their name and formality preference (formal vs. informal). The transition occurs once the Judge confirms these parameters are set or explicitly declined by the user.
    
    \item \textbf{State 2: EMOTION}
    \textit{Objective:} Identify the user's current emotional state.
    \textit{Logic:} The agent employs active listening to validate feelings. A strictly enforced constraint requires a minimum of two user messages in this state to prevent premature transitions, ensuring the user feels heard before moving forward.
    
    \item \textbf{State 3: EMOTION\_DECIDER (Internal Node)}
    \textit{Objective:} Route the conversation based on sentiment analysis.
    \textit{Logic:} An LLM classifier analyzes the dialogue history to categorize the user's state as \textit{Positive} or \textit{Negative}.
    \begin{itemize}
        \item \textbf{Negative:} Transitions to \textit{SUPER\_STATE\_EVENT} for processing triggers.
        \item \textbf{Positive:} Transitions directly to \textit{ASK\_EXERCISE}, skipping the event exploration phase.
    \end{itemize}
\end{itemize}

\subsubsection{Phase II: Exploration (Trigger Processing)}
\begin{itemize}
    \item \textbf{State 4: SUPER\_STATE\_EVENT}
    \textit{Objective:} Explore the antecedents of negative emotions.
    \textit{Logic:} Accessible only via the Negative path. The agent probes for specific events or triggers associated with the user's distress. Similar to the Emotion state, a minimum two-message constraint is enforced to ensure adequate exploration depth.
    
    \item \textbf{State 5: OPEN\_ENDED\_CONVERSATION}
    \textit{Objective:} Provide a safe space for deep elaboration.
    \textit{Logic:} If the user indicates a need to vent or elaborate further during State 4, the system transitions here. This state allows for free-form dialogue (minimum 4 messages) before gently steering the user back to the intervention track.
\end{itemize}

\subsubsection{Phase III: Therapeutic Intervention}
\begin{itemize}
    \item \textbf{State 6: ASK\_EXERCISE}
    \textit{Objective:} Obtain informed consent for the intervention.
    \textit{Logic:} The agent proposes a therapeutic exercise. The Response Retriever classifies the user's reply; affirmative responses lead to suggestion, while negative responses route to the closing phase.
    
    \item \textbf{State 7: EXERCISE\_SUGGESTION}
    \textit{Objective:} Deliver a context-aware SAT exercise.
    \textit{Logic:} This state utilizes the RAG system to select an exercise. The selection algorithm filters by the user's specific intervention day (such as Day 1: Exercises 1-3) and therapeutic stage (Beginning/Intermediate/Advanced), then uses the LLM to choose the specific activity that best aligns with the user's summarized emotional context.
    
    \item \textbf{State 8: EXERCISE\_EXPLANATION}
    \textit{Objective:} Guide the user through the selected practice.
    \textit{Logic:} The system provides step-by-step instructions. The user can request clarification, proceed to feedback, or request a different exercise (looping back to State 10).
    
    \item \textbf{State 9: FEEDBACK}
    \textit{Objective:} Debrief and consolidate learning.
    \textit{Logic:} The agent asks the user to reflect on their experience immediately post-exercise.
    
    \item \textbf{State 10: LIKE\_ANOTHER\_EXERCISE}
    \textit{Objective:} Iterative intervention loop.
    \textit{Logic:} The system offers additional exercises. An affirmative response loops the FSM back to \textit{EXERCISE\_SUGGESTION}, allowing for multi-exercise sessions; a negative response transitions to termination.
\end{itemize}

\subsubsection{Phase IV: Conclusion}
\begin{itemize}
    \item \textbf{State 11: THANKS}
    \textit{Objective:} Session closure and positive reinforcement.
    \textit{Logic:} The agent provides concluding remarks and encouragement.
    
    \item \textbf{State 12: END}
    \textit{Objective:} Terminal state.
    \textit{Logic:} The session is formally closed, and the final memory summary is committed to the long-term storage database.
\end{itemize}


\section{Screenshot of the Web-based Interface}
\label{appendix:interface}

\begin{figure}[H]
\centering
\includegraphics[width=1\linewidth]{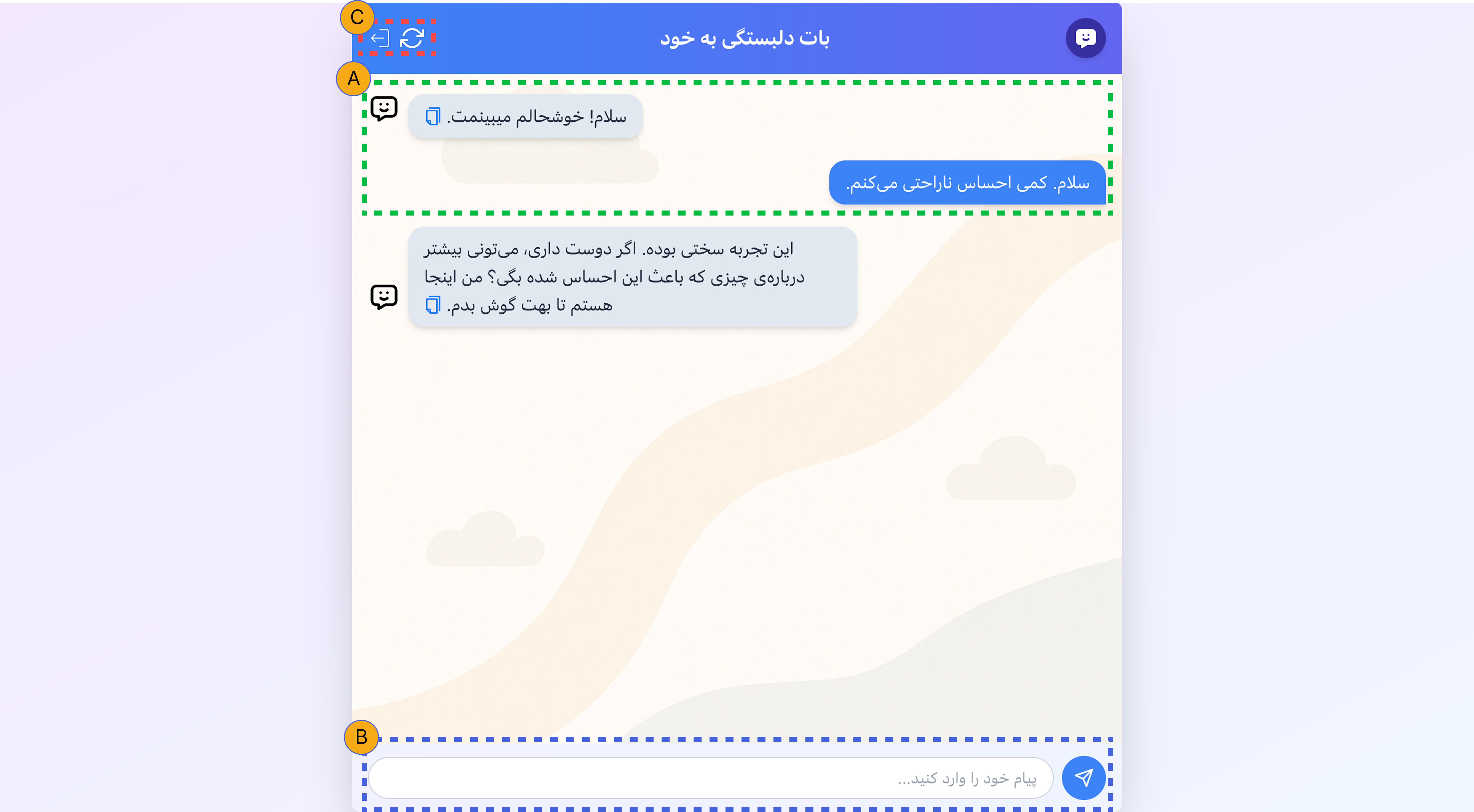}
\caption{Screenshot of the web-based user interface of the chatbot. After logging in, users are directed to the home screen where they can start interacting with the chatbot. (A) shows the list of user messages and corresponding chatbot responses. (B) is the input area for composing and sending messages to the chatbot. (C) contains the logout button on the left and the button to restart the conversation on the right.}
\end{figure}

\end{document}